%% file: main.tex
\documentclass[a4paper,fleqn]{cas-dc}
\ExplSyntaxOn
\cs_gset:Npn \__first_footerline:
  { \group_begin: \small \sffamily \__short_authors: \group_end: }
\ExplSyntaxOff

\usepackage[numbers]{natbib}
\usepackage{etoolbox}

\usepackage{graphicx}
\usepackage{textcomp}
\usepackage{xcolor}
\usepackage{float} 
\usepackage{dirtytalk}
\usepackage{cryptocode}
\usepackage{fancyvrb}
\usepackage{subcaption}
\usepackage{color}
\usepackage{wasysym}
\usepackage{hyperref}
\usepackage{adjustbox}
\usepackage{booktabs}
\usepackage{tabularx}
\usepackage{comment}
\usepackage{caption}
\hypersetup {
    colorlinks=true,
    linkcolor=blue,
    filecolor=blue,      
    urlcolor=blue,
    citecolor=blue
}

\def\tsc#1{\csdef{#1}{\textsc{\lowercase{#1}}\xspace}}
\tsc{WGM}
\tsc{QE}
\tsc{EP}
\tsc{PMS}
\tsc{BEC}
\tsc{DE}

\begin{document}
\sloppy
\let\WriteBookmarks\relax
\def\floatpagepagefraction{1}
\def\textpagefraction{.001}
\shorttitle{BlockMeter: An Application Agnostic Performance Measurement Framework For Private Blockchain Platforms}
\shortauthors{Ifteher Alom et~al.}


\title [mode = title]{BlockMeter: An Application Agnostic Performance Measurement Framework For Private Blockchain Platforms}

\tnotetext[1]{Corresponding Author: M. S. Ferdous}


\author[1]{Ifteher Alom}[
                        ]
\ead{ifteher.alom@gmail.com}

\author[2,3]{Md Sadek Ferdous}[
                        orcid=0000-0002-8361-4870
                        ]
\ead{sadek.ferdous@bracu.ac.bd}
\cormark[2]

\author[4]{Mohammad Jabed Morshed Chowdhury}[
                        orcid=0000-0003-4476-8882
                        ]
\ead{m.chowdhury@latrobe.edu.au}
\cormark[4]

\address[1]{Shahjalal University of Science and Technology, Sylhet, Bangladesh}
\address[2]{BRAC University, Dhaka, Bangladesh}
\address[3]{Imperial College London, London, UK}
\address[4]{La Trobe University, Melbourne, Australia}


\input{abstract}

\maketitle

\input{introduction}

\input{background}

\input{relatedwork}

\input{architecture}

\input{evaluation}

\input{experiment}

\input{discussion}
\input{conclusion}

\section*{Acknowledgement}
This article is the results of the research
  project funded by the AWS Cloud Credits for Research Program.



\bibliographystyle{unsrt}
\bibliography{main}




\end{document}

%% file: abstract.tex
\begin{abstract}
Blockchain Technology is an emerging technology with the potential to disrupt a number of application domains. Though blockchain platforms like Bitcoin and Ethereum have seen immense success and acceptability, their nature of being public and anonymous make them unsuitable for many enterprise level use-cases. To address this issue, Linux Foundation has started an open source umbrella initiative, known as the \textit{Hyperledger Platforms}. Under this initiative, a number of private blockchain platforms have been developed which can be used for different enterprise level applications. However, the scalability and performance of these private blockchains must be examined to understand their suitability for different use-cases. Recent researches and projects on performance benchmarking for private blockchain systems are very specific to use-cases and are generally tied to a blockchain platform. In this article, we present \textit{BlockMeter}, an application agnostic performance benchmarking framework for private blockchain platforms. This framework can be utilised to measure the key performance matrices of any application deployed on top of an external private blockchain application in real-time. In this article, we present the architecture of the framework and discuss its different implementation aspects. Then, to showcase the applicability of the framework, we use BlockMeter to evaluate the two most widely used Hyperledger platforms, Hyperledger Fabric and Hyperledger Sawtooth, against a number of use-cases.

\end{abstract}



\begin{keywords}
Blockchain \sep 
distributed consensus \sep
scalability \sep
performance \sep
platform agnostic \sep
performance benchmarking \sep
performance matrices
\end{keywords}

%% file: introduction.tex
\section{Introduction}
\label{sec_intro}
The concept of Blockchain Technology is undeniably an ingenious invention of Satoshi Nakamoto with the introduction of Bitcoin \cite{nakamoto2008bitcoin}. Equipped with a number of valuable properties \cite{FERDOUS2021103035, chowdhury2019comparative}, Blockchain technology has the potential to be one of the frontier technologies in the very near future. To investigate its potentiality, it is being actively investigated, within the academia, industries and the Governments around the world, how blockchain technology can be integrated with the existing use-cases or even how it can be leveraged to disrupt the traditional application domains, such as banking, finance, e-government, healthcare industries, IoT, agriculture and so on \cite{chowdhury2020survey, ferdous2019integrated}. Many industries in these application domains are investigating to shift their infrastructures towards blockchain. 

To cater to this need, in addition to Bitcoin, a number of new blockchain platforms, such as Ethereum \cite{ethereum2018}, Cardano, Algorand \cite{algoranddocs}, Polkadot \cite{Polkadotdocs} and others have emerged. These blockchain platforms are equipped with novel features such as smart contract and offer additional advantages in terms of scalability and performance in comparison to Bitcoin. However, these platforms are public in nature, meaning anyone can verify every single block, even observe every single activity and submit transactions anonymously or pseudonymously which make them unsuitable for enterprise level applications where privacy and identity are key requirements. In addition, there are issues with respect to scalability and performance of these blockchain platforms for any large-scale adoption in real-life settings \cite{dinh2017blockbench, pongnumkul2017performance}. 




To address these issues, Linux Foundation \cite{linuxfoundationhyperledger}, a non-profit technology consortium, has started an open source umbrella initiative, known as the \textit{Hyperledger Platforms}. The goal of this initiative is to achieve a industry-wide collaboration for developing enterprise-grade blockchain platforms. Under this initiative, a number of private Blockchain platforms have been developed which can be used for different applications. However, to utilise and select a particular blockchain platform for a specific use-case, one must be able to compare the available platform against a set of criteria and select the best suitable one.

One of the key factors that can help in the decision making of whether or not we should adopt a particular blockchain platform for a given application is the performance of the selected platform. There are a number of criteria that can be used to compare different blockchain platforms, however, the latency and throughput are often regarded as key matrices for any blockchain based application. 

Recently, there have a been a number of research and projects on performance benchmarking of private blockchain systems \cite{thesis01, dinh2017blockbench, pongnumkul2017performance,thakkar2018performance}. However, these projects are specific to use-cases and are generally tied to a blockchain platform. This means that the method presented in those works cannot be used to evaluate the suitability of any application for a specific blockchain platform. Furthermore, many of the experiments presented in those works have been conducted in simulated environments. Hyperledger Project itself has a tool called Caliper \cite{caliperhome}, unfortunately, it cannot measure the performance of blockchain applications deployed externally. To mitigate these limitations, in this article, we present \textit{BlockMeter},  an application agnostic, real-time performance benchmarking framework for private blockchain platforms. This framework can be utilised to measure the key performance matrices of any application deployed on top of an external private blockchain application in real-time.

\vspace{1mm}
\noindent \textbf{Contributions:} The major contributions of this article are presented below:

\begin{itemize}
    \item We analyse different performance metrics for private blockchain platforms.
    \item We present the architecture of BlockMeter, an application agnostic blockchain performance platform. 
    \item We discuss different implementation aspects of BlockMeter and describe how it can be integrated with different use-cases.
    \item Finally, to showcase the applicability of BlockMeter, we utilise BlockMeter to evaluate the two most widely used Hyperledger platforms, Hyperledger Fabric and Hyperledger Sawtooth, against a number of use-cases and blockchain network configurations.
\end{itemize}

\vspace{1mm}
\noindent \textbf{Structure:} In Section \ref{sec_background}, we briefly discuss about blockchain and its different aspects and explain different terminologies used in our research and the blockchain platforms on which we have conducted our experiments. Section \ref{sec_related_work} discusses briefly about some of the relevant research works within the scope of this article. In Section \ref{sec_arch}, we present the high level architecture of BlockMeter and discuss its implementation methodologies and protocol flows. In Section \ref{sec_experiments}, we present the conducted experiments and their corresponding results and in Section \ref{sec_comp} we compare different types of applications and their performance with respect to two blockchain platforms. We discuss different aspects related to BlockMeter in Section \ref{sec_discuss}. Finally, in Section \ref{sec_conclusion}, we conclude our article with a hint of future goals and expectations.


%% file: background.tex
\section{Background}
\label{sec_background}
Blockchain Technology is a promising invention that can be used to create immutable ledgers for a wide range of applications in a number of areas such as finance, banking, government, supply chains, academia and business enterprises \cite{chowdhury2020survey,ferdous2019integrated, alom2021dynamic}. Its development and applications are only in their emerging stages and a lot of efforts are being put by researchers in the academia and industries to explore how this technology can be exploited in practical use-cases in different application domains.

In this section some of the relevant key concepts are discussed. In the following, we briefly describe about blockchain (Section \ref{sec_blockchain_tech}), the Hyperledger projects (Section \ref{sec_hyperledegr}), Hyperledger Fabric (Section \ref{sec_fabric}), Hyperledger Sawtooth (Section \ref{sec_sawtooth}) and Hyperledger Caliper (Section \ref{sec_caliper}).

\subsection{Blockchain}
\label{sec_blockchain_tech}
In the year 2009 a digital currency named Bitcoin was introduced using Blockchain technology. First published in the previous year in a white paper authored by someone with the presumed pseudonym Satoshi Nakamoto \cite{nakamoto2019bitcoin}, Bitcoin promised the ability to carry out financial transactions without relying on a central authority (e.g. government backed central bank), utilising different cryptographic mechanisms. All transactions are kept on a ledger or the blockchain, which can be publicly accessed, ensuring transparency \cite{bitcoinhistory1}. 

Since then, Bitcoin has become the most widely-used decentralised digital currency. Its main technological breakthrough is due to its underlying mechanism called \textit{blockchain}, which is a smartly engineered decentralised system featuring an immutable append-only ledger of transactions shared and validated by a set of distributed network of Peer-to-Peer (P2P) nodes \cite{chowdhury2019comparative}. The ledger is essentially an ordered data structure consisting of many blocks of data linked together by cryptographic protocols. Each block contains some transactions where each transaction is a record of an action undertaken by a user to transact a certain amount of currency/data to another user/users. Each block refers to its previous block using a cryptographic hash, which refers to its previous block and so on, hence forming a chain which is colloquially known as \textit{blockchain}.

A smart-contract is a computer program which is deployed and executed on top of a computing platform underpinned by a blockchain. Such smart-contracts can be deployed using the notion of transactions. Additionally, transactions are used to invoke different functionalities of a smart-contract which ultimately changes the state of the blockchain. Being underpinned by a blockchain platform ensures that a smart-contract facilitates code immutability a sought-after feature in many application domains.

Despite the vast number of blockchain platforms and implementations available at this moment, we can broadly classify them into two types: permissionless (public) and permissioned (private) \cite{FERDOUS2021103035}.

A public blockchain allows any user to create a personal address and begin interacting with the network, by submitting transactions, and ultimately, adding entries to the ledger. Additionally, all users also have the opportunity to contribute as a node to the network, employing the verification protocols to help verify transactions and reach a distributed consensus regarding the state of the blockchain. On the other hand, a private blockchain acts as closed ecosystems, where only authorised users are able to join the network, see the recorded history, or issue transactions. Indeed, such blockchains are governed by specific members of a consortium or organisation and only the approved members and computer entities have the possibility of running nodes on the network, validating transaction blocks, issuing transactions, executing smart-contracts or reading the transaction history \cite{privpubdlt}. Bitcoin \cite{bitcoin2018} and Ethereum (Main-net) \cite{ethereum2018} are examples of public blockchains, whereas Hyperledger platforms \cite{hyperledger2018} and Quoram \cite{quorum2018} are examples of private blockchain systems.

As the popularity of Bitcoin, Ethereum and a few other derivative technologies grew, interests in applying the underlying technology of the blockchain and distributed ledger to more innovative enterprise use cases also grew. However, many enterprise use cases require performance characteristics that the currently available public blockchain platforms are unable to deliver. In addition, in many use cases, the identity of the participants is a hard requirement, such as in the case of financial transactions where Know-Your-Customer (KYC) and Anti-Money Laundering (AML) regulations must be followed \cite{fabricdocs}. In other words, in order to utilise blockchain technology in many enterprise use cases, we need to consider the following basic requirements \cite{fabricdocs}: (1) Participants must be identified/identifiable, (2) High transaction throughput performance, (3) Low latency of transaction confirmation and (4) Privacy and confidentiality of transactions and data. 

Therefore, in case of enterprise/business applications and use cases, mostly permissioned blockchains are utilised to restrict its access within the members of an organisation or consortium. 


\subsection{Hyperledger}
\label{sec_hyperledegr}
Hyperledger is an open source collaborative project undertaken to advance cross-industry blockchain technologies \cite{hyperledgersite}. It is a global collaboration, hosted by The Linux Foundation, which includes leaders from finance, banking, Internet of Things, supply chains, manufacturing and Technology. The Hyperledger project aims to accelerate industry-wide collaborations for developing high performance and reliable blockchain and distributed ledger based tools, standards and guidelines that could be used across various industry sectors to enhance the efficiency, performance of various business processes \cite{hyperledgersite2}. The various projects under Hyperledger umbrella include the following:
\begin{itemize}
    \item Hyperledger Fabric \cite{fabricdocs} - an enterprise-grade, permissioned blockchain platform for building various blockchain-based products, solutions and applications for business use-cases.
    \item Hyperledger Burrow \cite{burrowdocs} - a permissioned blockchain node that handles transactions and smart-contract on the Ethereum Virtual Machine, providing transaction finality and high transaction throughput on a proof-of-stake Tendermint consensus engine.
    \item Hyperledger Sawtooth \cite{sawtoothdocs} - an enterprise-level, permissioned, modular blockchain platform for building blockchain applications and networks using an innovative Proof of Elapsed Time \cite{poet01} consensus algorithm.
    \item Hyperledger Iroha \cite{irohadocs} - an easy to incorporate, efficient and trustworthy byzantine fault-tolerant tool having essential functionalities for assets, information and identity management.
    \item Hyperledger Composer \cite{composerdocs} - a set of tools that allows users to easily build, test and operate their own blockchain application. Unfortunately, currently the project is dysfunctional.
    \item Hyperledger Caliper \cite{caliperhome} - a blockchain benchmark tool that is used to evaluate the performance of a specific blockchain implementation.

\end{itemize}

\subsection{Hyperledger Fabric}
\label{sec_fabric}
Hyperledger Fabric \cite{fabrichomewebsite} is one of the projects under the Hyperledger umbrella. It is a permissioned blockchain framework, sometimes referred as a distributed operating system \cite{androulaki2018hyperledger} because of its highly modular and configurable architecture, enabling innovations, versatilities and optimisations for a broad range of industry use cases including banking, finance, insurance, healthcare, human resources, supply chain and even digital music delivery. Being permissioned means that the participants within the same blockchain network are known to each other, rather than anonymous and the trust requirement is dependant upon the application developed on top of it. This means that while the participants may not fully trust one another (they may, for example, be competitors in the same industry), a network can be operated under a governance model where a level of trust can be established because of activating a legal agreement or framework for handling disputes \cite{fabricdocs}. 


One crucial feature of Fabric is the support of general purpose programming languages for deploying smart-contracts, known as \textit{chaincode} in Fabric terminology. Currently, Fabric supports several languages such as Golang, JavaScript, Java and others \cite{fabricwiki}. Each Fabric chaincode has two public functions, the  \textit{Init} function which is called when the chaincode is deployed and the \textit{Invoke} function which is called every time the chaincode is called to access the ledger. A Hyperledger Fabric has different types of nodes. The committing peers maintain the ledger by holding a copy of it, while the endorsing peers are responsible for saving a copy of the ledger as well as receiving requests from the clients and validating them against the chaincode and endorsement policies (rules that govern who is allowed to do what within the blockchain platform). Finally, the orderers are responsible for batching and maintaining the order of transactions over the entire network. Ordering service may be single or multiple (by using Apache Kafka \cite{kafka01}. Since multiple organisations may participate in the same Fabric network, the Membership Service Provider (MSP) provides each peer with cryptographic identities and each organisation have a Certification Authority (CA) \cite{thesis01}. Hyperledger Fabric incorporates a new concept called channels that enables confidential transactions. A channel in Fabric is an instance of the blockchain with its own chaincode that can be accessed only by a predefined subset of participants.

\subsection{Hyperledger Sawtooth}
\label{sec_sawtooth}
Hyperledger Sawtooth \cite{sawtoothdocs} is another blockchain platform under the Hyperledger ecosystem. It has been designed to be highly scalable and efficient in terms of performance. Sawtooth supports different consensus protocols \cite{sawtooth126} like Practical Byzantine Fault Tolerance (PBFT) and Proof of Elapsed Time (PoET) which depends on Intel Software Guard Extencsion (Intel SGX). Intel SGX is a Trusted Execution Environment (TEE) built into the newer generation of Intel processors. SGX allows processing of instructions within a secure enclave inside the processor \cite{consensus01}.

The architecture of Sawtooth consists of four main components: validator, transaction processor, consensus engine and a REST API. The validator node is the heart of the system that is responsible for handling transactions, generating and storing blocks. The transaction processor node holds the business logic for validating transactions. The consensus engine uses an algorithm for ordering the transactions into blocks. The REST API node receives transactions from the clients and communicates with the validator nodes for getting the responses.

\subsection{Hyperledger Caliper}
\label{sec_caliper}
Hyperledger Caliper \cite{caliperhome} is a performance benchmark tool under the Hyperledger umbrella. It provides a generic method for the performance evaluation of several blockchain technologies. Caliper has a layered architecture that allows the separation of benchmark parameters from the ledger implementation. 

The generic benchmark layer monitors the resource consumption, evaluates the recorded data and generates a report at the end of the test. Blockchain interfaces allow different blockchain implementations to communicate with the benchmark layer. The adaptation layer consists necessary modules for different ledger implementations \cite{thesis01}. The adaptation layer is used to integrate any existing blockchain system into Caliper framework using network configuration files. The corresponding adapter implements the 'Caliper Blockchain NBIs (North Bound Interfaces)` by using corresponding blockchain's native SDK. 

The interface and Core layers implement core functions and provide NBIs. Four kinds of NBIs are provided \cite{caliperdocs}:
\begin{itemize}
    \item Blockchain operating interfaces performs operations such as deploying smart-contracts, invoking contracts and querying states from the ledger.
    \item Resource Monitor performs operations to start/stop a monitor and fetch resource consumption status of any backend blockchain system, including CPU, memory, network IO and others.
    \item Two kinds of monitors are provided - one is to watch local/remote docker containers, and another is to watch local processes. 
    \item Performance Analyzer: It contains operations to read predefined performance statistics and print benchmark results. Key metrics are recorded while invoking blockchain NBIs. Those metrics are used later to generate the statistics. 
    \item Report Generator: It contains operations to generate an HTML format test report.
\end{itemize}

The major interaction with Caliper takes place via two configuration files. The first one is a ledger specific configuration file that consists the network details like roles, chaincode, policies and others. The second file is used to specify parameters regarding the tests and workloads that are to be carried out like labels and number of tests, number of clients, transaction rate and workload callback. The benchmarking criteria in Caliper so far include success rate, transaction throughput, transaction latency and resource consumption. However, the main limitation of using Hyperledger Caliper for benchmarking or for analysing the performance of a blockchain application is that we cannot execute the test operations on a externally deployed blockchain platform. Caliper requires the platform specifications to be fed into its architecture and it deploys a blockchain platform accordingly. However. in real life blockchain platform may be deployed in a complex system environment (for example using docker containers, using virtual machines connected remotely, over a local organisation network and so on) which cannot be replicated in Caliper. In short Caliper performs its operations in a closed and controlled environment which leaves a gap between the observed performance and reality. In addition, Caliper only tests the performance of the underlying blockchain platform, completely detaching the overlay application and thus, it is difficult to benchmark a fully integrated application using Caliper. Because of all these factors, it is difficult to create an application agnostic benchmarking framework for real-life deployments using Caliber. 

\subsection{Performance metrics}
\label{sec_analysis}

There are a number of parameters which can be utilised to compare and analyse the performance of blockchain systems. From our research and references form existing works as discussed in Section \ref{sec_related_work}, we have identified a number of such parameters that are mostly used in private blockchain platforms and these parameters are: throughput, latency and resource consumption. Next, we present a brief discussion of these parameters.
\begin{itemize}
    \item \textbf{Throughput:} Throughput is defined as the number of transactions that are successfully completed and committed to the ledger in unit time. Throughput is determined by the capacity of the network. In other words, network infrastructure must be sufficient to handle the traffic. Theoretically, throughput will increase as the input load is increased, up to the the maximum network capacity \cite{herwanto2021measuring}. However, network capacity, apart from physical setup, also depends on the internal logic execution and access control mechanism. In terms of performance evaluation, high throughput is a key requirement in most of the applications at enterprise levels. It is generally expected that the system is capable of tackling maximum requests even under high traffic within a business environment.
    
    \item \textbf{Latency:} Latency can be defined as the time elapsed between a transaction submit and transaction completion. Latency is dependent on propagation delay, transit and queuing of the system. The protocol flow of the system plays a crucial role in either increasing or decreasing the latency. High latency is a major issue with the most of the established blockchain platforms like Bitcoin, because of its huge ledger \cite{yasaweerasinghelage2017predicting}. This is why enterprises are looking for private blockchain systems to avoid additional overheads. The system should respond within minimal delay under high traffic, keeping up with all the protocols at the same time.
    
    \item \textbf{Resource consumption:} Resource consumption is referred to as the extent to which resources such as CPU and main memory of the host system is used by different nodes of the blockchain network during its execution. A blockchain system requires a network of nodes which have different roles and perform various operations in order to maintain the integrity of the network and ledger. Despite the various benefits that a blockchain platform provides, one of the major subjects of discussion is its significant requirements for resources. Public blockchain systems such as Bitcoin and Ethereum are maintained by power and resource hungry mining rigs \cite{jurivcic2020optimizing}. This is another reason businesses are looking for private blockchain systems that can be efficient in utilising the resources \cite{blockchaintypesdiff}.
\end{itemize}

At the application level, the task that we are trying to solve may require one or more chaincode and specific configurations and may interact with the ledger at various frequencies. Moreover, the language and the framework of  the  chaincode  also might  have  some  impact  because  of  the  compilation and execution time. Having said that, these matrices are primarily the indicators of the performance of a system for a specific workload that is under test.

A Workload can be defined as the load that is executed on a blockchain platform to evaluate its performance and its ability to handle a specific use case. Elementary workloads aim to focus on a single aspect of the network.  These include \textit{DoNothing} (minimal chaincode functionality: read/write), \textit{CPUHeavy} (heavy computation chaincode) and \textit{DataHeavy} (chaincode involving large size data and parameters). These generic workloads are simulated implementations of realistic use cases. For example, Fabcar (a demo application supplied with the Fabric codebase), that performs read, write and insert operations on a predefined set of records using key value pairs can be taken in the category of a DoNothing chaincode.

Apart from the matrices discussed above, factors like network topology, distribution of roles and resources allocated to the nodes, validation and endorsement policies also affect the network performance.  At the ledger level the distribution of chaincode among the peers, storage (database) type,  consensus mechanisms and block-size have significant impact in the performance of the ledger \cite{thakkar2018performance}.

%% file: relatedwork.tex
\section{Related Work}
\label{sec_related_work}

The performance analysis of private blockchain platforms is a relatively new dimension of research, however, there are already a few works in this domain. In this section, we briefly review and analyse the existing works.

Leppelsack et al. have described a generic implementation of a framework that can measure the performance of Hyperledger Fabric under different workloads using different chaincode \cite{thesis01}. To simulate the workload Hyperledger Caliper has been used. Caliper deploys the network, installs the chaincode and submits transactions at a predefined rate and reports the latency and resource consumption of different nodes. Four different experiments have been conducted using a Fabric network consisting of a single orderer and two organisations, each having two peers. The impact of varying transaction rate, chaincode, block-size and network loss have been studied graphically.

The experiments show that the ledger system builds up a backlog after the transaction rate exceeds 300 tps (Transaction per Second). The \textit{DoNothing} chaincode results in a delayed backlog while others have backlog built up earlier. The \textit{DataHeavy} chaincode creates a slight latency for a workload of 10KB while the \textit{CPUHeavy} chaincode has a significant latency. A reduced block-size improved the performance for higher transaction rates. While the reduction of 10 transaction per block to 5 resulted in doubling the latency at high transaction rates. For a low transaction rate of 100 tps a network loss of 5\% resulted in almost 60\% increase in transaction latency.

It is clear that transaction rate, workload, block-size and network loss all together have significant impact on the performance of Hyperledger Fabric, however, none of them alone can be considered as a bottleneck. Some of these factors like the block-size can be tuned and optimised based on the use case, while some factors like network loss may not be controllable.



The authors in \cite{dinh2017blockbench}, have  demonstrated the  development and analysis of the first private blockchain evaluation framework - Blockbench. The authors have used the framework to study the performance of three different private blockchains- Ethereum, Parity and Hyperledger Fabric under different workloads. According to their experimental results, the current state and performance of private blockchain systems are not mature enough to replace the existing database systems. Blockbench framework allows integration of various blockchain platforms via simple Application Programming Interfaces (API) which is also a feature of our project. Different kinds of chaincode have been used to simulate various use cases. The measurement matrices include throughput, latency, resource consumption of the nodes, scalability. The evaluation mainly focuses on comparison of different blockchain implementation based on the matrices. The impact of ledger specific configurations (e. g. block-size) has not been considered.

The authors in \cite{pongnumkul2017performance}, have presented an evaluation of Hyperledger Fabric and Ethereum. For both of the blockchain technologies they have implemented an application with different smart contracts. The execution times of the smart contracts are recorded and compared. However, the main focus lies in the performance of these blockchain platforms by varying the amount of transactions that are executed. Metrics on the ledger layer including execution time, transaction latency and transaction throughput have been inspected. These metrics are evaluated for both platforms with varying numbers of transactions. Overall this allows to compare the performance of different workloads as well as making statements on the pros and cons of the two technologies.

The experiments show the differences in execution time of varying number of transactions, with different platforms and different functions. The execution time increases as the number of transactions in the data set increases. The  execution time of Hyperledger Fabric is consistently lower than Ethereum and the gap between the execution time of Fabric and Ethereum also grows
larger as the number of transactions increase.

As the number of transactions in the data set grows, the latency of transactions in Ethereum worsens considerably more compared to Hyperledger Fabric. It has been also found that Fabric has higher throughput than Ethereum in all of the experimental data sets. However, for similar computational resources, Ethereum can handle more concurrent transactions.

The authors have suggested that estimating the expected number of transactions will be very crucial in selecting suitable platforms as they can alter subsequent throughput, execution time, and latency. Particularly, latency can play a crucial role in applications involving money transfer as well as other forms of trading.

The authors in \cite{thakkar2018performance} have conducted a comprehensive empirical study to understand and optimise the performance of Hyperledger Fabric, a permissioned blockchain platform, by varying configuration parameters such as block-size, endorsement policy, channels, resource allocation, and state database choices. The experiments have been conducted using a two-phased approach. In the first phase, the impacts of various Fabric configuration parameters such as block-size, endorsement policy, channels, resource allocation, state database choice on the transaction throughput and latency have been studied. It has been found that endorsement policy verification, sequential policy validation of transactions in a block, and state validation and commit (with CouchDB) have been the three major bottlenecks.

In the second phase, they have focused on optimising Hyperledger Fabric v1.0 based on the observations gained in the first phase. Several optimisations have been introduced and studied, such as aggressive caching for endorsement policy verification in the cryptography component, parallelising endorsement policy verification and bulk read/write optimisation for CouchDB during state validation.



%% file: architecture.tex
\section{Architecture, Implementation \& Protocol Flow}
\label{sec_arch}
The core theme of BlockMeter is that it advocates the concept of an application agnostic performance measurement framework for private blockchain systems. Based on the analysis as discussed above in section \ref{sec_analysis}, we have worked out a Proof of Concept (PoC), to evaluate our proposal and formulate some insights on the performance of two popular Hyperledger blockchains.

In this section, we present the architecture of BlockMeter (Section \ref{subsec:archi}) and the protocol flow (Section \ref{sec_flow}) which illustrates how different components of the architecture interact for a particular testing scenario. Based on the architecture, we have developed a Proof of Concept (PoC) in order to evaluate its applicability. The implementation details for the PoC are described in Section \ref{sec:implementation}.


\subsection{Architecture}
\label{subsec:archi}
\begin{figure}[h!]
    \centering
    \includegraphics[width=\columnwidth]{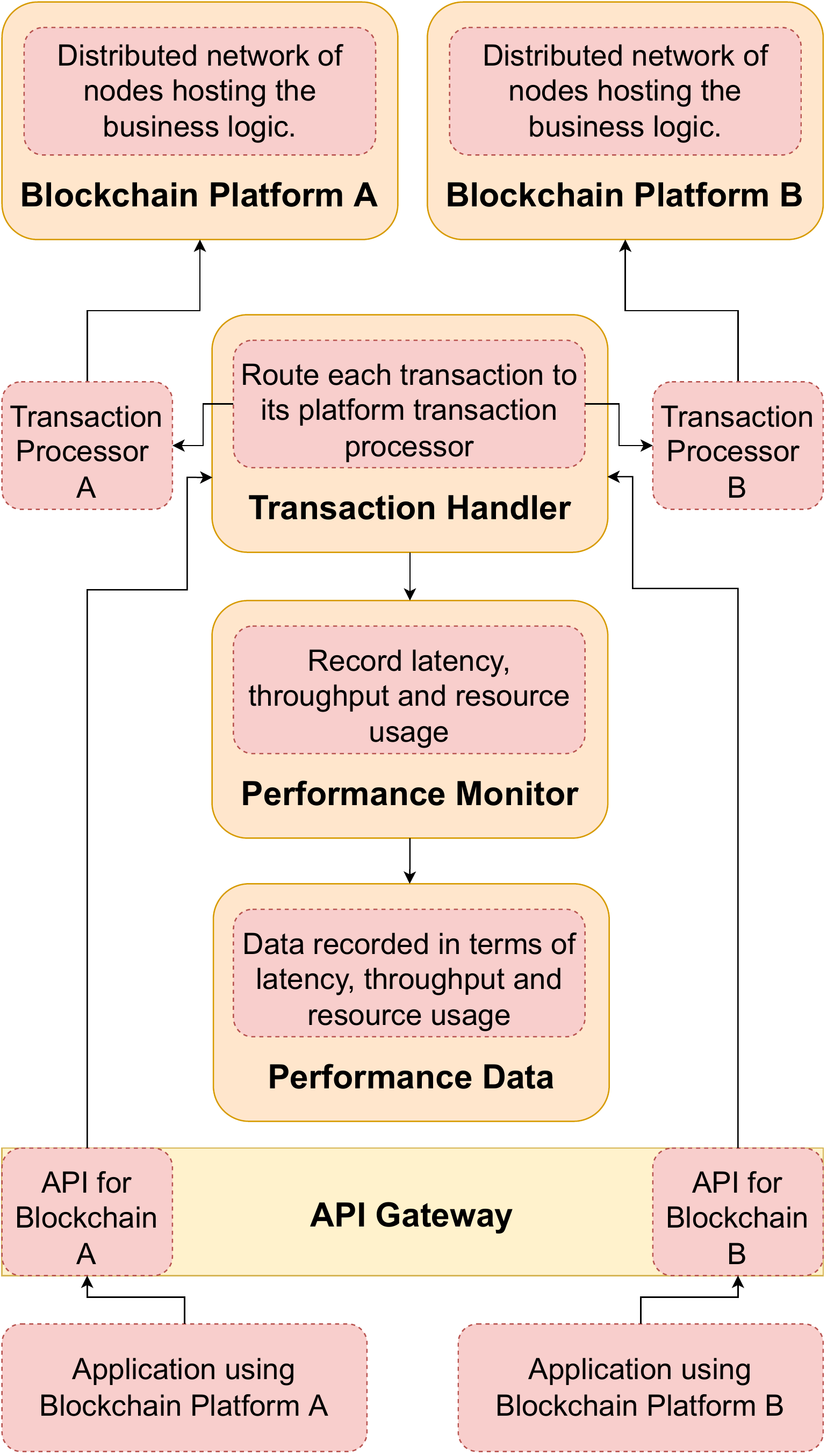}
    \caption{Architecture of BlockMeter}
    \label{arch_image}
\end{figure}

The architecture, as described in figure \ref{arch_image}, consists of several modules. Each module performs a specific task by receiving some data, applying a set of instructions on it and then forwarding it to the next module. Next, we elucidate the architecture and its different components.


\begin{itemize}
    \item Transactions Handler: It is one of the central components that receives incoming requests and initiates the process. Its main role is to translate incoming HTTP requests into blockchain platform specific data-frames. Every blockchain platform has its own prerequisites and conditional input fields that are used at various levels of the transaction processing, this module encapsulates the incoming request into a blockchain object. At the end of the process, when the transaction is completed, the object is parsed to get the necessary information and the client is notified with a response.
    
    \item Performance Monitor: This can be called the crux of the system as it is responsible for collecting the performance data as the transactions are being cascaded from the framework to the blockchain and vice versa. Several methods are embodied into this module to collect data like transaction start time, end time, and the status of the docker containers. The data thus collected are continually uploaded to specific files in the directory, which can be retrieved for further analysis.
    
    \item Transaction Processor: This component translates client requests into objects that can be perceived by a particular blockchain platform within its ecosystem. The respective Software Development Kit (SDK) for each blockchain platform provides the necessary functions and libraries to achieve this task. In our framework we have exploited the SDK to develop such a middleware for Hyperledger Fabric and Hyperledger Sawtooth platforms. More blockchain platforms can be incorporated here to make the support range wider and comprehensive.
    
    \item Blockchain Application This module comprises of the use-case or client application under test. It is responsible for setting up a blockchain based application and its constituent elements independently. The corresponding transaction processor of the blockchain platform performs the operations that has to be undertaken during the performance testing thereby making our proposed tool capable of monitoring the performance under real traffic and congestion of the surrounding environment.
    
    \item API Gateway: In order to serve requests from the client application each supported blockchain platform has been assigned a particular path where the client requests are received and relayed to the the transaction processor. Finally, upon the completion of the request the response is sent back to the client. 
    
    \item Performance Data: Upon completion of each round of load testing we receive the data recorded by the performance monitor. The data consists of request start times and end times, throughput count and system statistics from the docker containers that host the blockchain application. These data can be parsed and studied for further analysis based on the requirements and criteria of evaluation.
\end{itemize}




\subsection{Implementation}
In order to develop and implement the PoC, we would need to integrate different private blockchain platforms with the framework. From all the available private blockchain platforms, we have selected Hyperledger Fabric and Sawtooth from the Hyperledger Umbrella Projects \cite{hyperledgersite} for the PoC as these two are popular private blockchain platforms. The implementation instance of the developed PoC based on the proposed architecture of HyerpScale is illustrated in Figure 2.
\begin{figure}[h!]
    \centering
    \includegraphics[width=\columnwidth]{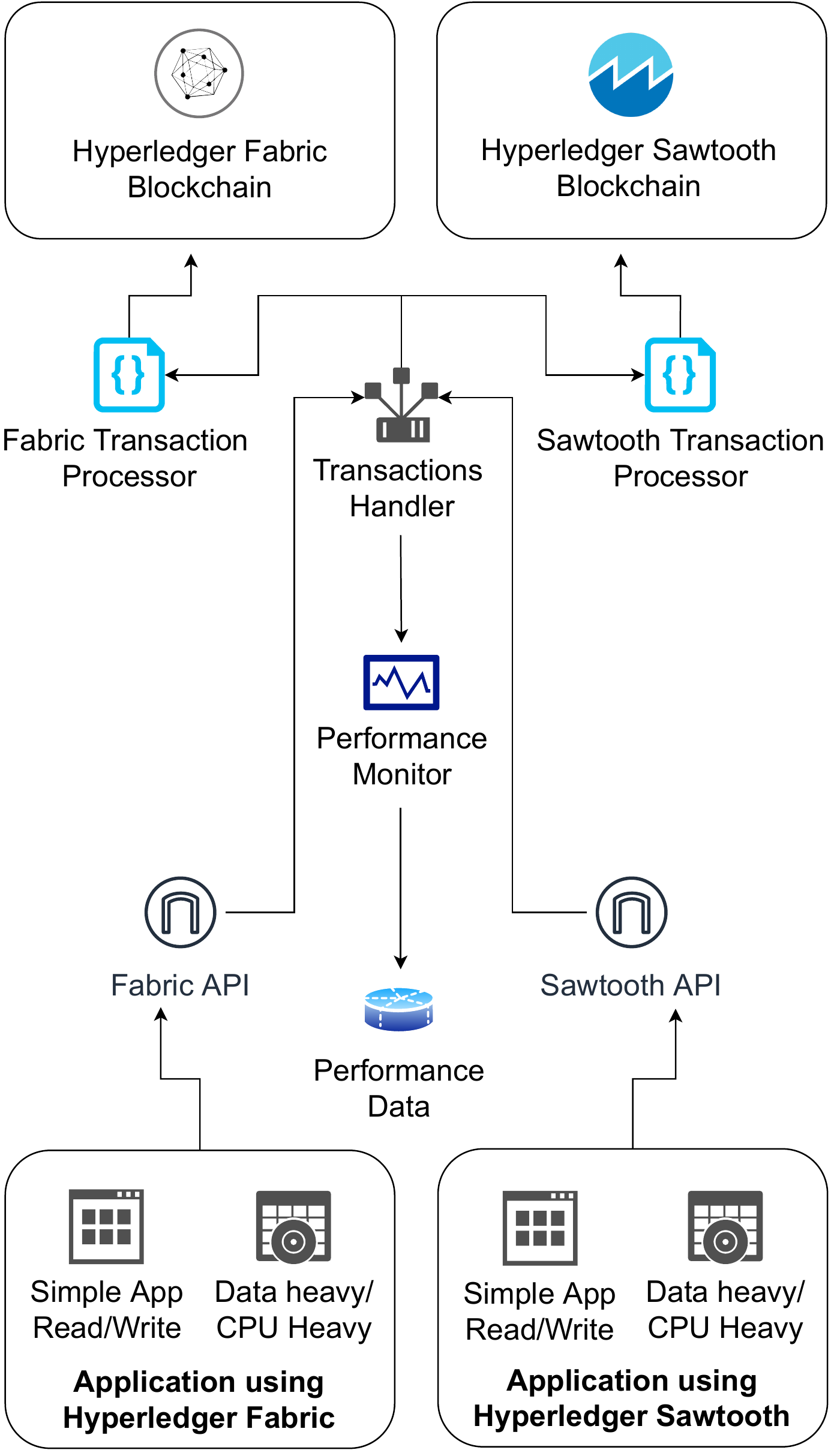}
    \caption{Reference Implementation of BlockMeter}
    \label{impl_image}
\end{figure}
\label{sec:implementation}

At the Blockchain Application level (Blockchain Application part of Section \ref{sec_arch} and in Figure \ref{arch_image}), we have we have configured two blockchain networks as discussed before. The Hyperledger Fabric network consists of a number of different types of nodes as discussed earlier. Similarly, for the Sawtooth network, required validating nodes and transaction processor nodes have been set up. For both cases, the blockchain application have been deployed using AWS EC2 instances and Docker (used for virtualisation). The network topology and structure have been varied for different experiments (discussed in Section \ref{sec_experiments}). 
    
The Target SDK for Hyperledger Fabric are available in Golang, Java, Python and NodeJS while Hyperledger Sawtooth SDK are available in python, NodeJS and Rust. We have used the NodeJS SDKs in our implementation to create a middleware (the Transaction Processor in section \ref{sec_arch}), which accepts  client requests and complies them into platform specific objects that can communicate with the backend blockchain network using the SDK functions as discussed in Section \ref{sec:implementation}

The Performance recorder is a server set up using NodeJS and Express \cite{express} (a web application framework for NodeJS). The server has been hosted on an AWS EC2 machine exposing two POST APIs. The requests are routed to the blockchain network at the backend using an instance of the appropriate class. This server also facilitates transaction throughput monitoring which tracks the number transactions being completed per second and resource monitoring which records the resource consumption of the peer nodes every three seconds (because each request for the docker statistics takes more than two seconds to respond).

Two API ends, acting as the Gateway APIs, are exposed from the NodeJS-Express server. A client application submits transaction requests to different APIs based on the blockchain platform that is under test. The request is converted into a transaction proposal and then submitted to the corresponding transaction handler which pre-processes the transaction for further processing as discussed in Section \ref{sec_flow}.

The Fabcar \cite{fabcar01} (a baseline barebone application provided in the Hyperledger Fabric repository which performs simple blockchain operations such as read, write and update) application has been used as the basis for the Blockchain Application that performs simple ledger operations like create, read and update. For the Sawtooth platform, we have developed a transaction processor similar to the Fabcar application, so that we can compare the results under similar testing scenarios. Several network configurations consisting of single and multiple ordering and validating nodes have been deployed using Docker so as to identify the variation in performance with respect to different network configurations.


\subsection{Protocol Flow}
\label{sec_flow}
\begin{figure*}[h]
    \centering
    \includegraphics[width=\textwidth]{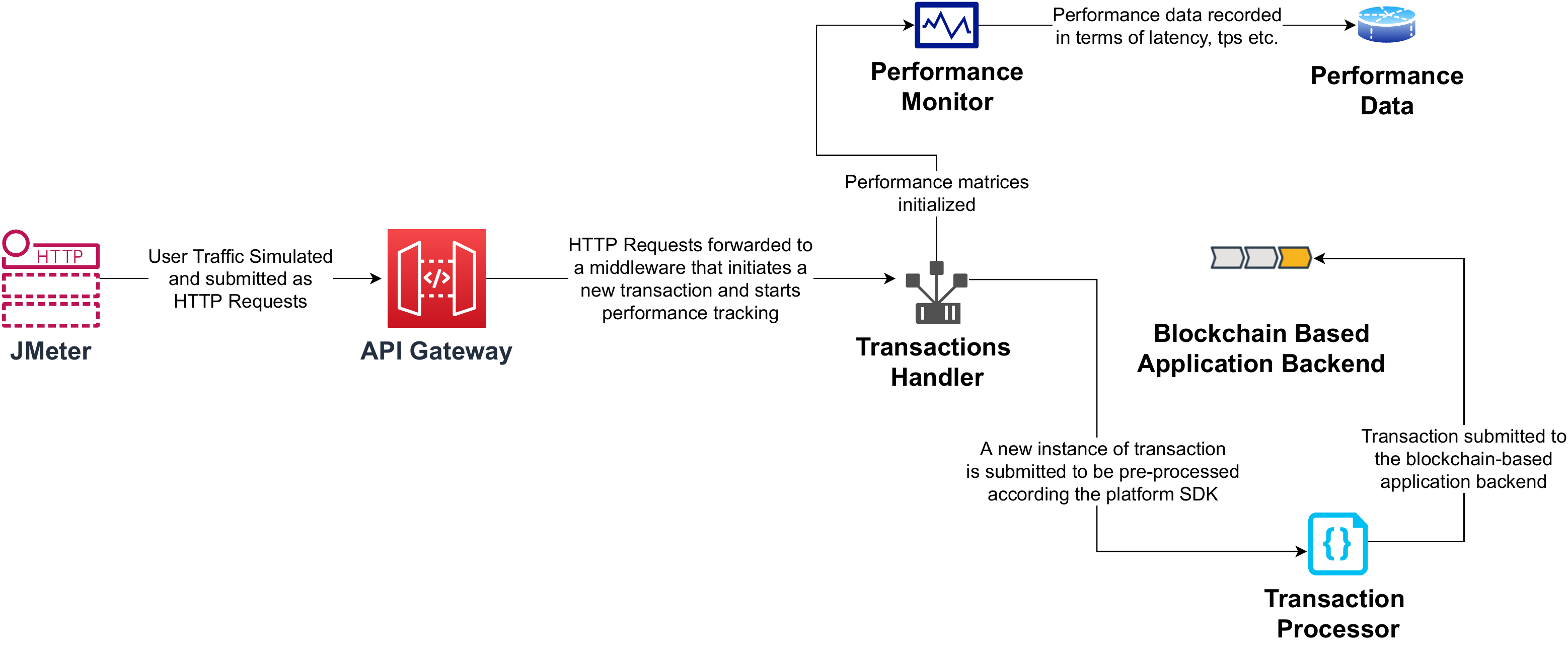}
    \vspace{-20mm}
    \caption{Protocol Flow}
    \label{fig:protocol_flow}
    \hspace*{\fill}
    \vspace{0mm}  
\end{figure*}

Now, we illustrate a protocol flow utilising the framework in Figure \ref{fig:protocol_flow}. This flow showcases the interactions between different components of the architecture when used in a testing scenario.

\begin{itemize}
    \item \textbf{User Creation}: A number of user accounts have been created for each blockchain platform. These user credentials consist of the user names and their respective public and private keys that are used to sign the transaction requests before submitting them to the blockchain platform. These accounts have been used to simulate scenarios where a number of users simultaneously  use the application built on top of that particular blockchain platform.
    
    \item \textbf{Payload Creation}: In order to perform an operation in the blockchain (i.e. invoking a particular function of the smart-contract in the blockchain) we need to provide the necessary parameters. In this step such parameters are created programmatically with random payload.
    
    \item \textbf{Traffic/Load Generation}: The payload and parameters generated earlier are imported into Apache JMeter \cite{jmeter01} that injects them into HTTP requests and hits the framework's Gateway API. JMeter is an open source software quality evaluation tool maintained by the Apache Software Foundation \cite{apachesoftfound}. It is primarily used for load testing, unit testing and  basic performance monitoring of web based applications. Using JMeter, the number of requests/load generated per second is varied over the span of each experiment.
    
    \item \textbf{Transaction Request}: As illustrated in Figure \ref{fig:protocol_flow},  the JMeter acts as the source to generate transaction requests. These requests are essentially HTTP requests consisting of transaction particulars for the respective blockchain platform. We can feed in different configuration parameters, e.g. the number of simulated users, the frequency of user requests and so on, from a pre-loaded CSV. JMeter submits these user requests to the \textit{API Gateway}. Upon the receipt of HTTP requests at the API gateway of the framework, an instance of each request is created and performance matrices are initialised attached to it. Each request is validated for valid parameters and the data is normalised into the required structure.
    
    \item \textbf{Request Pre-processing}: Once the transaction parameters are retrieved from the requests, the API gateway forwards these to the \textit{Transactions Handler} where an instance of the transaction is created and the restructured data object is sent to the corresponding transaction processor of the blockchain platform, where the request is pre-processed as per the platform requirements. This includes the creation of a signed transaction instance that consists of the payload and execution instructions.
    
    \item \textbf{Performance Monitoring}: Once a transaction instance is ready to be submitted to the blockchain, the performance monitor is notified with a signal to start recording performance matrices and the usage status of docker containers/instances that are hosting the blockchain network.  Finally, the \textit{Performance Data} consisting of transaction latency, transaction throughput and resource consumption are recorded in JSON format into the file system. The records are then normalised, reformatted and converted into tabular format using NodeJS \cite{nodejs} which are then visually analysed using python libraries.
    
    \item \textbf{Transaction Response}: After successfully passing through all protocols and validation policies of the blockchain network, a response is received at the framework. At this stage, performance matrices for the transaction instance is finalised and Apache JMeter is responded with a success status code.
\end{itemize}

%% file: evaluation.tex
\section{Experiments}
\label{sec_experiments}
To show the applicability of the developed performance measurement framework, we have designed and carried out a number of experiments. In this section we present the details on various aspects of the experiments conducted using the framework.

\subsection{Experimental Setup}
\label{sec:setup}
The principal motivation of the experiments is to showcase how the developed framework can be utilised to gauge the performance, in terms of throughput, latency and resource consumption, of different private blockchain platforms, Hyperledger Fabric and Sawtooth in this instance. 

Towards this aim, we have conducted several experiments and load testing by simulating some of the typical use cases under varying testing conditions (like low and high request traffic and number of operating nodes of the blockchain platform). To conduct these experiemnts, our system has been deployed in Amazon Web Services (AWS) EC2 instances powered by 4 VCPUs and 16 Gigabytes of RAM. The different blockchain network configurations have been initiated with various orderer nodes using docker in AWS EC2 instances as well. 

Every use case involves a streamlined flow of operations (refer to Protocol Flow in section \ref{sec_flow}) that is initiated/requested by the Apache JMeter in the experiment, followed by a set of pre-processing at the framework and finally, concluded by appending the transaction to a blockchain. Next, we present a brief overview of the courses of actions involved in this regard.

In order to demonstrate a comparison of the performance matrices between the chosen blockchain platforms, we have selected three different types of applications that are exemplary to typical user applications. 
\begin{itemize}
    \item \textbf{Simple Application:} A simple key-value store that stores certain values against unique IDs within the blockchain. We have used the Fabcar application as the simple application. This application performs operations such as insert, update and delete on the data. It is a representative of a typical read/write application.
    \item \textbf{Data Heavy Application:} This application handles large amount of data in every transaction. It performs read and write operation on records that are 10 Kilobytes in size. For our experiment we have used a modified version of the Fabcar application. It represents a use case of a data intensive application.

    \item \textbf{CPU Heavy Application:} In this application we have modified the chaincode to introduce a CPU intensive task consisting of repeated arithmetic operations on random numbers. Every transaction is constrained to perform the mathematical operation before it comes to completion. This is an example of a  use case of a typical application that involves intensive computation.
\end{itemize}

These three applications represent three broad categories of applications that may use blockchain in their backend data validation and processing. These findings will give us an insight on the performance of Hyperledger Fabric and Hyperledger Sawtooth when use cases require handling of some intense tasks.

In a nutshell, the procedure involved in a single round of an experiment began by importing randomly generated payload at Apache JMeter which was then translated into a transaction and submitted to the blockchain platform while tracing the performance in terms of latency, throughput and resource consumption at the same time. For different scenarios,  different  simultaneous  cycles, rate of incoming traffic or the load were varied steadily. For Hyperledger Fabric, experiments began at 10 requests per second that gradually was increased to 1000. For Hyperledger Sawtooth, we began at 10 transactions per second (tps),  but had to wind up at 600 tps (details discussed in section \ref{sec_discuss}). At the end of the experiments the data recorded by the performance monitor is represented using graphs to facilitate a comprehensive performance analysis of the underlying platform.

\begin{figure*}[h]
  \begin{subfigure}{0.49\textwidth}
    \includegraphics[width=\columnwidth]{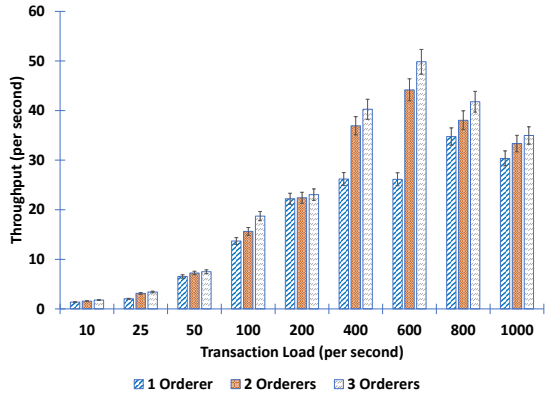}
    \caption{Throughput vs Load}  \label{fig:tpsfab}
  \end{subfigure}
  \hspace*{\fill} 
  \begin{subfigure}{0.49\textwidth}
    \includegraphics[width=\columnwidth]{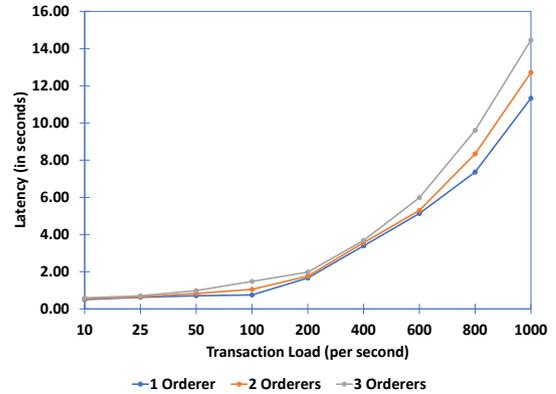}
    \caption{Latency vs Load}  \label{fig:latencyfab}
  \end{subfigure}
  \begin{subfigure}{0.49\textwidth}
    \includegraphics[width=\columnwidth]{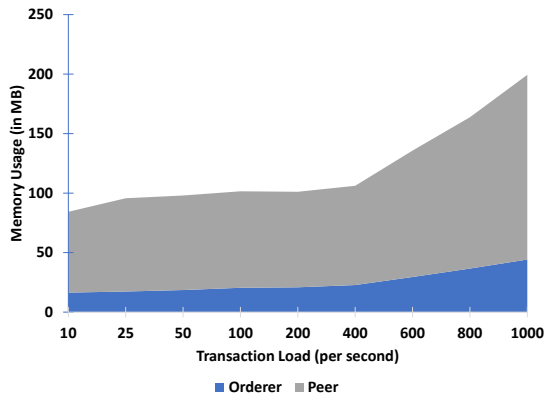}
    \caption{Memory usage: Orderer vs Peer}  \label{fig:fabmem1}
  \end{subfigure}
  \hspace*{\fill}
  \begin{subfigure}{0.49\textwidth}
    \includegraphics[width=\columnwidth]{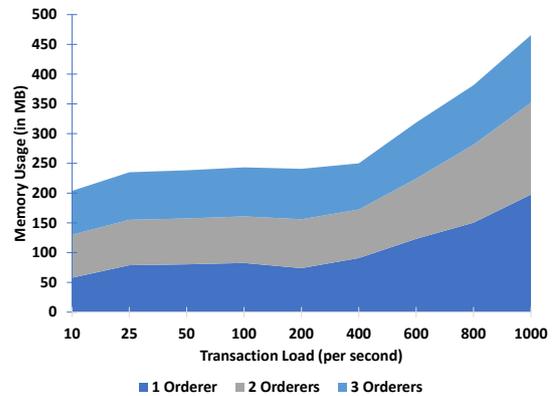}
    \caption{Memory usage by number of nodes}  \label{fig:fabmem2}
  \end{subfigure}
  \caption{Performance of Hyperledger Fabric} \label{fig:performfab}
\end{figure*}

\begin{figure*}[h]
  \begin{subfigure}{0.49\textwidth}
    \includegraphics[width=\columnwidth]{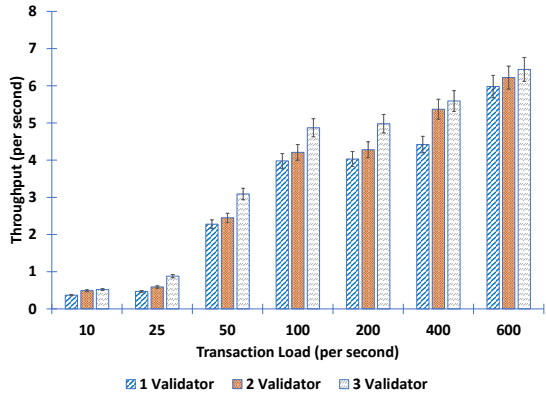}
    \caption{Throughput vs Load}  \label{fig:tpssaw}
  \end{subfigure}
  \hspace*{\fill} 
  \begin{subfigure}{0.49\textwidth}
    \includegraphics[width=\columnwidth]{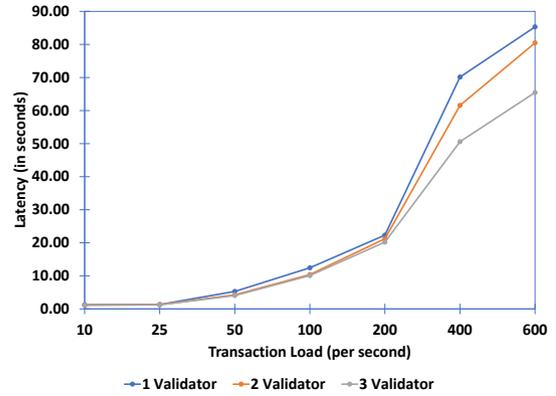}
    \caption{Latency vs Load}  \label{fig:latencysaw}
  \end{subfigure}
  \begin{subfigure}{0.49\textwidth}
    \includegraphics[width=\columnwidth]{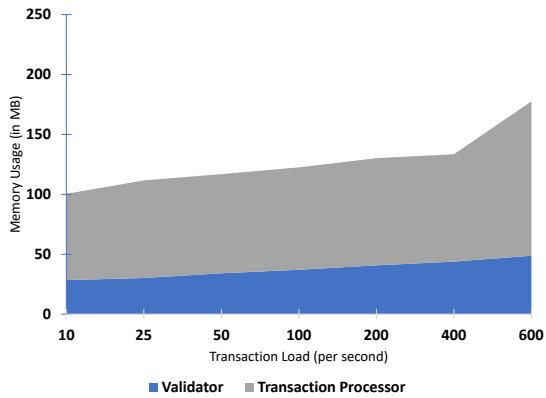}
    \caption{Memory usage: Validator vs Transaction Processor}  \label{fig:sawmem1}
  \end{subfigure}
  \hspace*{\fill} 
  \begin{subfigure}{0.49\textwidth}
    \includegraphics[width=\columnwidth]{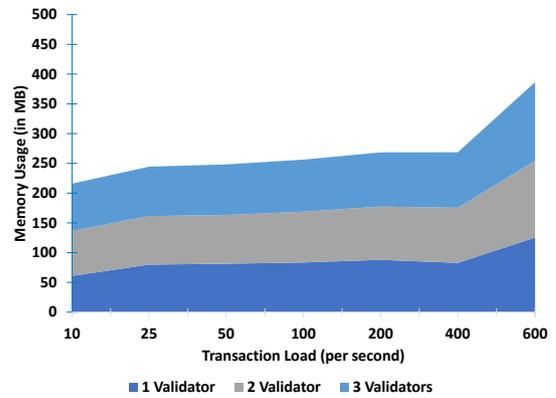}
    \caption{Memory usage by number of nodes}  \label{fig:samwmem2}
  \end{subfigure}
  \caption{Performance of Hyperledger Sawtooth} \label{fig:performsaw}
\end{figure*}

\subsection{Findings}
\label{sec_findings}


In this section, we explore the overall performance of Hyperledger Fabric and Hyperledger Sawtooth in terms of throughput, latency and resource consumption under varying numbers of operating nodes (orderers and peers in Fabric and validators and processors in Sawtooth). 

\vspace{3mm}
\noindent \textbf{Hyperledger Fabric}: The performance results of the experiments involving Hyperledger Fabric have been plotted in Figure \ref{fig:performfab}. An important observation from Figure \ref{fig:tpsfab} is that the transaction throughput of Fabric has increased steadily with increasing traffic. However, the trend levelled off and began to decrease when the input traffic load crossed 600 per second. The increase in the number of operating nodes (ordering service and peers) did result in an increase of the throughput slightly for every single transaction load group. The primary reason for such behaviour is that a typical blockchain transaction has to filter through a number of security policies and requirements before it is appended to the main blockchain. This intricate secure mechanism of the blockchain system incurs a delay and backlog which results in a comparatively lesser throughput under higher traffic. 

The latency in the transaction response increases significantly as the number of transactions arriving per second increases (Figure \ref{fig:latencyfab}). The situation did not improve even when multiple ordering service and peer nodes are configured into the network. The time in seconds clearly demonstrates that we need more optimisation and improvement in the architecture of the Fabric platform. The complex architecture and protocol flow of Fabric system can be accounted for the delayed response. Moreover, more operational nodes means more synchronising overhead which may overrun the added computation. However, there are several tweaks that we can make: like setting the block-size and block creation timeout that may improve the response time depending upon the situation and type of application \cite{thakkar2018performance}. 

The memory usage increases as the number of transactions arriving per second is increases as seen in Figure \ref{fig:fabmem1}. In Fabric, peer nodes consume more memory as they perform significant amount of tasks for validation as well as store a copy of the blockchain. As a consequence its memory consumption is higher and keeps growing as more transactions are appended. Multiple ordering nodes further increases the memory usage as seen in Figure \ref{fig:fabmem2}. However, the numbers in megabytes are not significant given the availability of memory in modern computing devices.


\vspace{3mm}
\noindent \textbf{Hyperledger Sawtooth}: It can be observed from Figure \ref{fig:tpssaw} that transaction throughput has increased gradually with increasing traffic. Also, multiple validating nodes lead to a slight improvement in the throughput performance. Hyperledger Sawtooth uses a hardware dependent consensus algorithm known as Proof of Elapsed Time (PoET) that relies upon Software Guard Entension (SGX) capability of the processor \cite{sgxintel}. Intel SGX is a new type of Trusted Execution Environment (TEE) integrated into the new generation of Intel processors. SGX enables the execution of code within a secure enclave
inside the processor, whose validity can be verified using a remote attestation process supported by the SGX \cite{ferdous2020blockchain}. The working principle of PoET algorithm involves a leader election, waiting time, and SGX facility. As a result the performance is poorer in comparison to Fabric which uses Ordering Service and endorsement policies for consensus. 

In terms of transaction latency of Sawtooth, the graph in Figure \ref{fig:latencysaw} shows that the response time increased significantly as the transaction load is increased. With the addition of multiple validating nodes and transaction processors the delay is improved slightly. However, the results reveal that Sawtooth is way behind Hyperedger Fabric when it comes to response latency. As highlighted earlier, the consensus algorithm of Sawtooth and the underlying mechanism that is reliant on SGX facility and leader election is the key reason for such delayed transaction processing. More importantly, this excessive backlog has accounted for almost a system freeze when inbound traffic reached 800 requests per second and higher.

With respect to the the memory usage, Sawtooth shows a higher rate of memory consumption than Fabric. Figure \ref{fig:sawmem1} illustrates that the memory usage increases almost steadily as the transaction arrival rate is increased. Multiple validator nodes further increase the resource availability and usage (Figure \ref{fig:samwmem2}). In Sawtooth, transaction processors consumer higher portion of the memory as it performs most of the validation and stores a copy of the blockchain.

%% file: experiment.tex
\section{Comparison of Different Applications}
\label{sec_comp}

\begin{figure*}[h]
  \begin{subfigure}{0.49\textwidth}
    \includegraphics[width=\columnwidth]{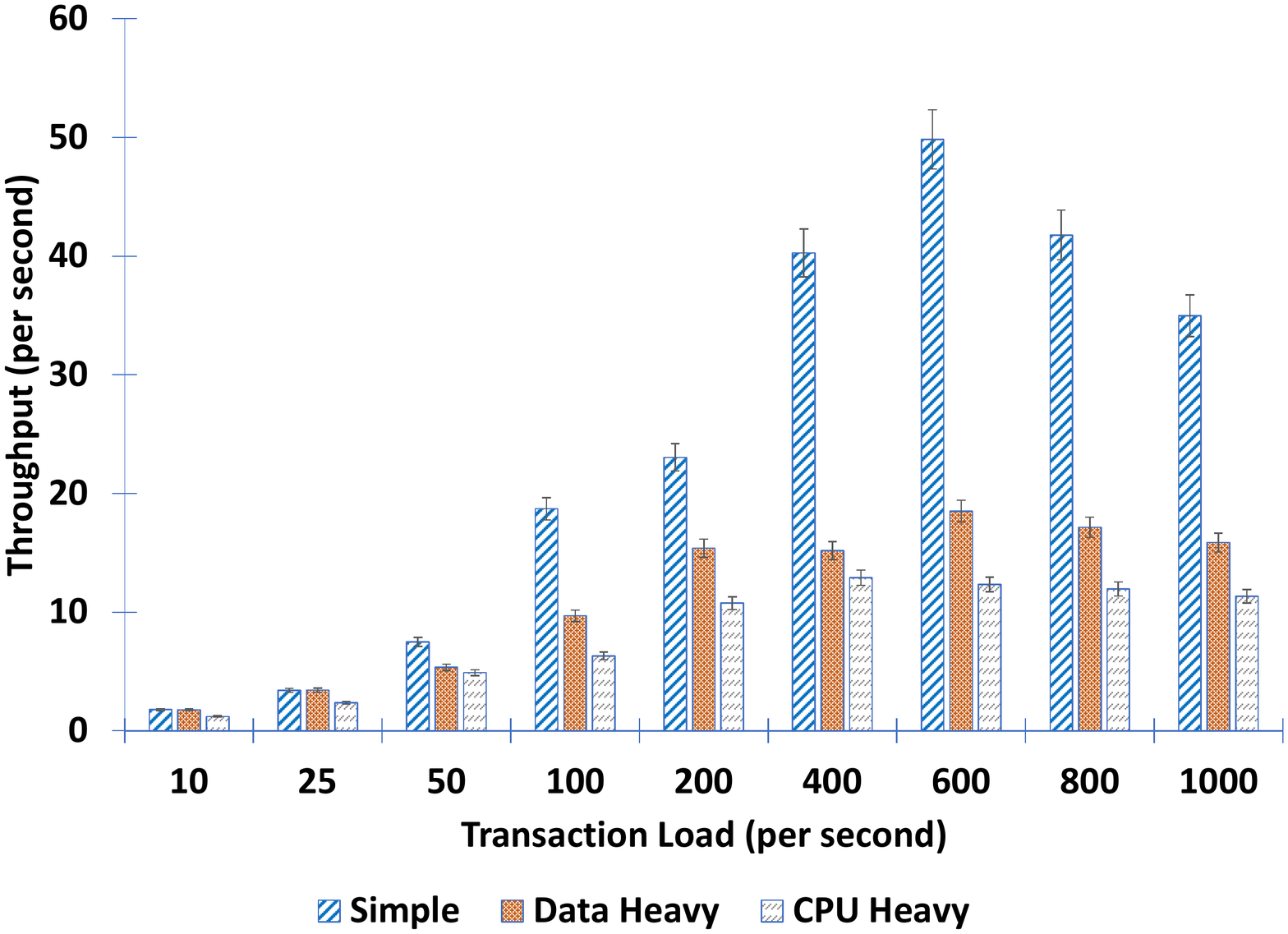}
    \caption{Throughput vs Load}  \label{fig:tpsfabcomp}
  \end{subfigure}
  \hspace*{\fill} 
  \begin{subfigure}{0.49\textwidth}
    \includegraphics[width=\columnwidth]{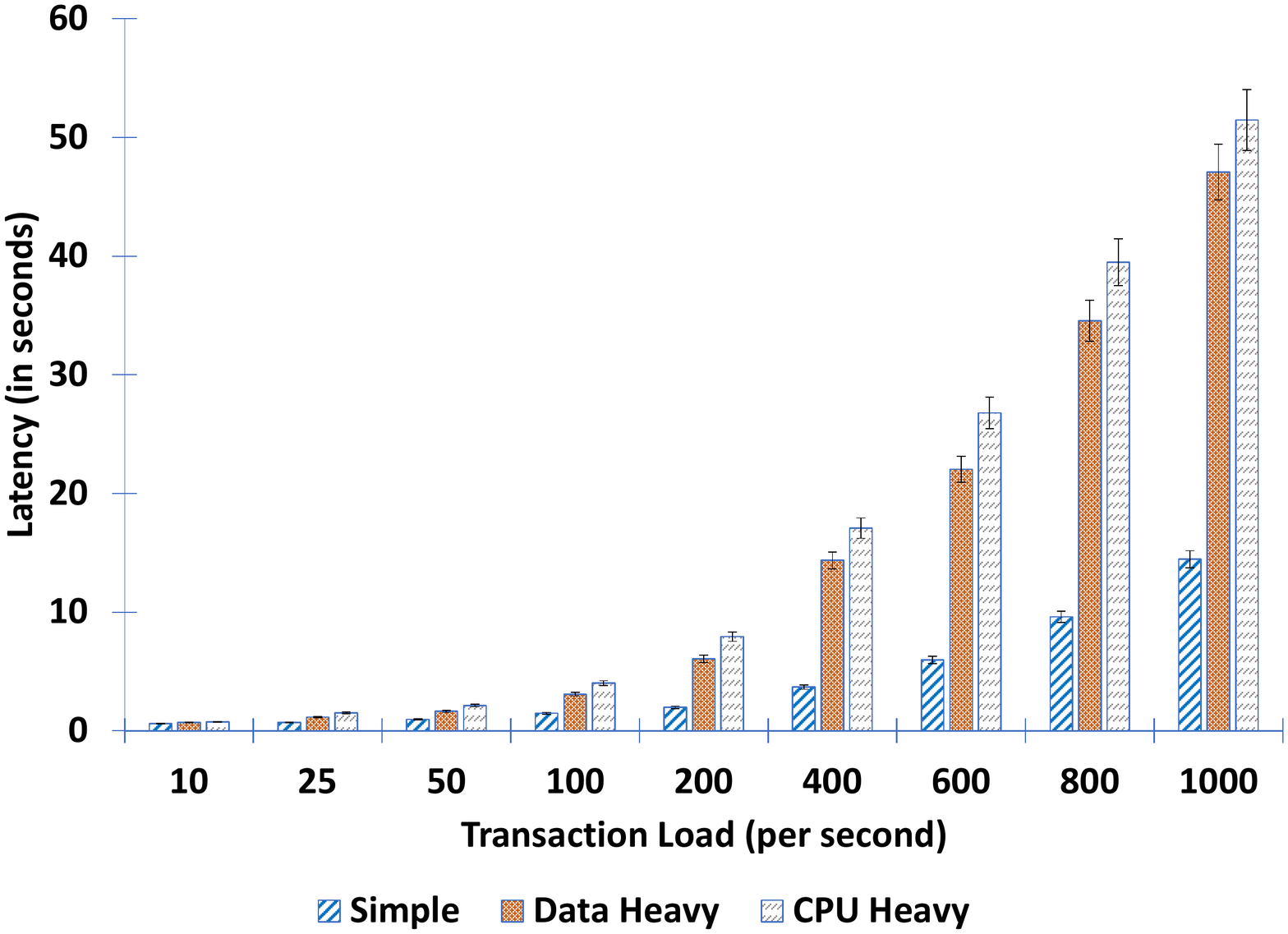}
    \caption{Latency vs Load}  \label{fig:latfabcomp}
  \end{subfigure}
  \caption{Hyperledger Fabric Application Performance} \label{fig:fabcomp}
\end{figure*}
\begin{figure*}[h]
  \begin{subfigure}{0.49\textwidth}
    \includegraphics[width=\columnwidth]{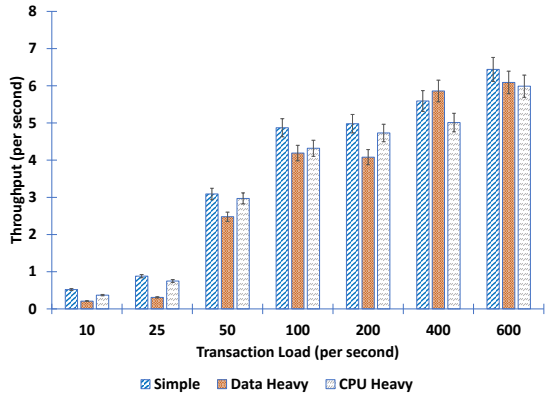}
    \caption{Throughput vs Load}  \label{fig:tpssawcomp}
  \end{subfigure}
  \hspace*{\fill} 
  \begin{subfigure}{0.49\textwidth}
    \includegraphics[width=\columnwidth]{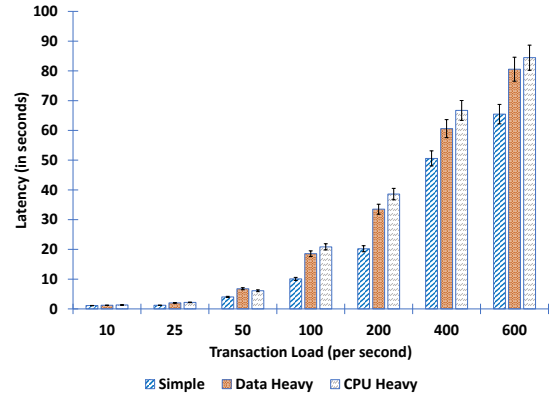}
    \caption{Latency vs Load}  \label{fig:latsawcomp}
  \end{subfigure}
  \caption{Hyperledger Sawtooth Application Performance} \label{fig:sawcomp}
\end{figure*}

The comparative results of the application centric experiments for Fabric and Sawtooth are plotted in Figure \ref{fig:fabcomp} and Figure \ref{fig:sawcomp} respectively.

With respect to the throughput of Hyperledger Fabric (Figure \ref{fig:tpsfabcomp}), \textit{Simple} application provides the maximum throughput as expected. The \textit{CPU Heavy} application gives the minimum throughput and \textit{Data Heavy} application shows a slightly better output. As observed earlier in Section \ref{sec_findings}, we can observe the throughput falling when input traffic exceeds 600 transactions per second. As expected, in case of latency (Figure \ref{fig:latfabcomp}), \textit{CPU Heavy} application has the highest response delay, followed by \textit{Data Heavy} application and \textit{simple} application. The \textit{CPU Heavy} application being overwhelmed with intensive computation and the \textit{Data Heavy} application loaded with large size parameters, such behaviours are understandable. What is striking though is that these numbers might deter enterprise application development where quicker response may be a key requirement.

On the other hand, Hyperledger Sawtooth has shown a much deteriorating result in case of both throughput and latency (Figure \ref{fig:sawcomp}). Although throughput has increased gradually with increasing traffic (Figure \ref{fig:tpssawcomp}), the latency has increased drastically (Figure \ref{fig:latsawcomp}). It can almost certainly be said that applications at production environment require a much better output for these areas of performance. It is observed that \textit{Simple}, \textit{Data Heavy} and \textit{CPU Heavy} with increasing payload size and logical complexity respectively, accounted for a slightly poorer performance. As discussed in Section \ref{sec_findings}, the consensus algorithm of Sawtooth platform is a major reason for this poor performance. It goes without saying Sawtooth blockchain framework has a lot of scopes for improvement in comparison to Fabric.

%% file: discussion.tex
\section{Discussion}
\label{sec_discuss}


As blockchain technology and its areas of applications in the mainstream are being heavily contemplated, it is crucial to have a methodology to efficiently measure its practicality in terms of performance under various conditions. In the course of our research we have realised that performance measurement and analysis in the domain of blockchain, have been primarily limited to public blockchain platforms like Bitcoin \cite{bitcoin2018} and Ethereum \cite{ethereum2018}. Very few research works have considered private blockchain platforms and in them they have compared private blockchain platforms with popular public blockchains, which is not very fruitful as the two systems function quite differently and are intended to dissimilar audiences.

This article presents BlockMeter, an application agnostic performance measurement framework for private blockchain platforms. The application agnostic feature stems from the idea that it is not coupled with any particular application. Indeed, any blockchain application can utilise the framework to measure and compare the performance of different private blockchain platforms for that particular application by submitting/simulating user interactions via any load testing software such as Apache JMeter. This has been possible as our framework exposes different APIs which act as the entry point for outside applications to interact with the system. The PoC that we have developed currently measures the performance of Hyperledger Fabric and Hyperledger Sawtooth.

From the different related work and analysis, it has been found that the latency and throughput are the two key decisive elements of performance in the field of blockchain applications. As such, we have focused more on these matrices to determine the performance of the applications in our experiments. With the advent of silicon chips, memory and computing capability have sky rocketed in recent times. In addition, power and energy consumption are directly linked to resource usability and efficiency of a computer system. Hence, we have also incorporated a resource usage recording mechanism within BlockMeter, allowing us to consider resource usage as a performance determiner. That is why the throughput, latency and resource consumption have been included as measurement matrices with our framework.  

\vspace{3mm}
\noindent \textbf{Advantages:} To summarise, BlockMeter has the following advantages:
\begin{itemize}
    \item BlockMeter is capable of measuring performance matrices like latency, throughput and resource usage of an independently deployed blockchain application. The recorded matrices are key indicators of the system's performance and can act as determining factors of its applicability for a particular business application.
    \item The performance framework can be easily integrated with applications developed using Hyperledger Fabric and Hyperledger Sawtooth platforms, The application specific configuration are minimal as the main execution system (i,e the blockchain and its configuration) is independent from the framework.
    \item Our tests have been conducted by hosting BlockMeter and deploying the blockchain applications on Amazon AWS  EC2 \cite{awsec2} machines. This has helped us evaluate the performance in a more realistic environment in contrast to a controlled local test setting.
    \item BlockMeter has been designed in a modular fashion that allows more blockchain platforms to be easily integrated into it. 
\end{itemize}



\begin{table*}[h]
\centering
\input{comp-table}
\captionsetup{justification=centering}
\caption{Comparative analysis of existing works with BlockMeter}
\label{tab:tabrw}
\end{table*}

\vspace{3mm}
\noindent \textbf{Comparative analysis:} Next, we analyse how BlockMeter fares in comparison to other relevant works discussed in Section \ref{sec_related_work}. The comparative analysis is summarised in Table \ref{tab:tabrw} against a number of criteria. Table \ref{tab:caliper-comp-table} highlights the key points that give BlockMeter an upper hand over Caliper \cite{caliperdocs}. In the tables, the symbol `\CIRCLE' has been used to denote a certain criteria is satisfied by the respective work whereas the symbol `\Circle' indicates that a certain criteria is not satisfied.

As it can be seen from table \ref{tab:tabrw}, BlockMeter uses an API based approach for ensuring that it is not tied to a particular application, thus facilitating its application agnostic feature. In contrast, the top level applications are tightly coupled in most of the works. This means, significant changes need to be made whenever a new application is integrated within their systems. Also, BlockMeter provides a modular framework, giving it the flexibility to add additional blockchain platforms easily. This is also absent in most of the works.

The evaluation matrices provide the key insights from a performance evaluation framework. Most of the related researches have focused only on latency and throughput for  their analysis while others also considered resource consumption. However,  BlockMeter is the only one which considered all three evaluation matrices. BlockMeter also considered three different types of applications as per with many works. BlockMeter is also equipped with a performance monitoring mechanism that makes it self sufficient for recording performance data, whereas most of the related work had to rely on Caliper \cite{caliperdocs} or some other tools. 

Also, most of the researches have been conducted in local servers or controlled environments which eliminates many overhead that arises in real life environments. For example, some works deployed their implementations within local servers. The BlockMeter framework and the blockchain platforms have been deployed in AWS EC2 \cite{awsec2} virtual machines via web service settings which allowed us to perform the tests in a more realistic setting. Moreover, some of the research have used Hyperledger Caliper \cite{caliperdiagram} or general load testing tools for the evaluation which limits the scope of experiment as the target blockchain is deployed internally by the testing tool itself, rather than a realistic and independent externally deployed platform.


\begin{table}[h]
\centering
\input{caliper-comp-table}
\captionsetup{justification=centering}
\caption{Caliper vs BlockMeter}
\label{tab:caliper-comp-table}
\end{table}

Next, we present a comparative analysis between Caliper and BlockMeter. The result of the analysis is summarised in Table \ref{tab:caliper-comp-table} where the symbols denote the semantics discussed previously. From the table, it can be observed that BlockMeter provides a number of advantages and benefits in comparison to Caliper \cite{caliperdocs}. To begin with, Caliper is designed with a much tightly-coupled layers of components that limit its flexibility to be modified and configured to specific needs for a particular application. On the contrary, BlockMeter provides a very modular architecture that enables anyone to easily configure it for any application. Furthermore, BlockMeter provides an API gateway for any application to interact with the underlying blockchain framework, thus facilitating an application agnostic performance evaluation. Also, unlike Caliper, the components of BlockMeter are designed to support easier integration with other blockchain platforms beyond the Hyperledger project. However, BlockMeter and Caliper have certain common aspects as well. Both are capable of monitoring similar performance matrices (latency, throughput and resource usage) and generate tabular summary based on the recorded experiment data.

\vspace{3mm}
\noindent \textbf{Limitations:} Our developed BlockMeter framework is capable of measuring crucial performance matrices of private blockchain platforms, nevertheless, it has some limitations.

\begin{itemize}
    \item The current implementation is limited to only two private blockchain platforms, Hyperledger Fabric and Hyperledger Sawtooth. There are other private blockchain platforms emerging as well. However, we have designed the framework in such a way that other blockchains with similar architectural setup can be easily integrated and we have plans for full-fledged support for other Hyperledger projects \cite{hyperledgersite2}.
    
    \item Our existing implementation does not provide a central dashboard with the summary of the executed tests. An automated visual interface would be very helpful for any end user to get primary insights of the test results.
    
    \item Our implementation is generally focused on comparison and analysis of private blockchain platforms with are more favourable for enterprise applications. However, public blockchains are also being used by a huge number of users. We would also like to explore the possibility of adding a sub-module for crypto-currency based blockchains to study and analyse their performance.
\end{itemize}

%% file: comp-table.tex
\centering
\begin{tabular}{l|l|c|c|c|c|c} 
\toprule
\multicolumn{2}{l|}{Related Work}             &\begin{tabular}[c]{@{}c@{}}~H. F. Leppelsack\\ \cite{thesis01} \end{tabular} 
& 
\begin{tabular}[c]{@{}c@{}}Blockbench\\ \cite{dinh2017blockbench} \end{tabular} & \begin{tabular}[c]{@{}c@{}}S. Pongnumkul, et al.\\ \cite{pongnumkul2017performance} \end{tabular} & \begin{tabular}[c]{@{}c@{}}P. Thakkar et al.\\ \cite{thakkar2018performance} \end{tabular} & \begin{tabular}[c]{@{}c@{}}BlockMeter\\ \end{tabular} \\
\midrule
\multirow{4}{*}{\begin{tabular}[c]{@{}l@{}}Architecture \\Design\end{tabular}}
& API Based & \Circle  & \CIRCLE  & \Circle  & \Circle  & \CIRCLE\\ \cline{2-7}
& Modular  & \Circle  & \CIRCLE  & \Circle & \Circle & \CIRCLE \\ \cline{2-7}
& Framework  & \Circle  & \CIRCLE  & \Circle & \Circle & \CIRCLE \\ \cline{2-7}
& App. Agnostic  & \Circle  & \Circle  & \Circle  & \Circle  &\CIRCLE \\
\midrule
\multirow{2}{*}{\begin{tabular}[c]{@{}l@{}}Blockchain \\Platforms\end{tabular}} & Fabric & \CIRCLE & \CIRCLE & \CIRCLE & \CIRCLE & \CIRCLE\\ \cline{2-7}
& Sawtooth  & \Circle & \Circle & \Circle & \Circle & \CIRCLE \\ 
\midrule
\multirow{2}{*}{\begin{tabular}[c]{@{}l@{}}Evaluation \\Tool\end{tabular}}
&  Caliper & \CIRCLE & \Circle & \Circle & \Circle & \Circle \\ \cline{2-7}

& Other & \Circle & \CIRCLE & \CIRCLE & \CIRCLE & \CIRCLE \\ 
\midrule
\multirow{3}{*}{\begin{tabular}[c]{@{}l@{}}Evaluation \\Matrices\end{tabular}}  & Latency & \CIRCLE & \CIRCLE & \Circle & \Circle & \CIRCLE \\ 
\cline{2-7}

& Throughput & \CIRCLE & \CIRCLE & \CIRCLE & \CIRCLE & \CIRCLE\\ \cline{2-7}

& Resource & \Circle & \Circle & \CIRCLE & \CIRCLE & \CIRCLE \\ 
\midrule
\multirow{4}{*}{\begin{tabular}[c]{@{}l@{}}Application \\Type\end{tabular}}  & Read/Write & \CIRCLE & \CIRCLE & \CIRCLE & \CIRCLE & \CIRCLE \\ \cline{2-7}
                                                        
& Data Heavy & \CIRCLE & \CIRCLE & \Circle & \Circle & \CIRCLE \\ \cline{2-7}

& CPU Heavy & \CIRCLE & \CIRCLE& \Circle & \Circle & \CIRCLE \\
\midrule
\multirow{2}{*}{\begin{tabular}[c]{@{}l@{}}Testing \\Environment\end{tabular}}  
& Web Service & \Circle & \Circle & \CIRCLE & \Circle & \CIRCLE \\\cline{2-7}

& Local Server & \CIRCLE & \CIRCLE & \Circle & \CIRCLE & \Circle \\
\bottomrule
\end{tabular}

%% file: caliper-comp-table.tex
\centering
\begin{tabular}{l|c|c} 
\toprule
Feature                                                                                                            & Caliper \cite{caliperdocs}                & BlockMeter  \\ 
\midrule
\multicolumn{1}{l|}{\begin{tabular}[c]{@{}l@{}}Application agnostic\\performance evaluation\end{tabular}}          & \multicolumn{1}{c|}{\Circle} & \CIRCLE           \\ 
\midrule
\multicolumn{1}{l|}{\begin{tabular}[c]{@{}l@{}}API based load injection\\and application testing\end{tabular}}     & \multicolumn{1}{c|}{\Circle} & \CIRCLE           \\ 
\midrule
\multicolumn{1}{l|}{\begin{tabular}[c]{@{}l@{}}Loosely coupled modular\\architecture\end{tabular}}                 & \multicolumn{1}{c|}{\Circle} & \CIRCLE           \\ 
\midrule
\multicolumn{1}{l|}{\begin{tabular}[c]{@{}l@{}}Scope for integrating other\\blockchain platforms\end{tabular}}     & \multicolumn{1}{c|}{\Circle} & \CIRCLE           \\ 
\midrule
\multicolumn{1}{l|}{\begin{tabular}[c]{@{}l@{}}Quick set-up and usability\\with external application\end{tabular}} & \multicolumn{1}{c|}{\Circle} & \CIRCLE           \\ 
\midrule
\multicolumn{1}{l|}{\begin{tabular}[c]{@{}l@{}}Latency, throughput and\\resource monitoring\end{tabular}}     & \multicolumn{1}{c|}{\CIRCLE} & \CIRCLE           \\ 
\midrule
\multicolumn{1}{l|}{\begin{tabular}[c]{@{}l@{}}Experimental data summary\end{tabular}}     & \multicolumn{1}{c|}{\CIRCLE} & \CIRCLE           \\  
\bottomrule           
\end{tabular}

%% file: conclusion.tex
\section{Conclusion}
\label{sec_conclusion}

Blockchain is a promising technology that aims to solve many of the security and integrity issues of our traditional data processing systems. However, this elegantly engineered system is still under development phases. In recent times, a great deal of interest has been observed within the academia and industry which is a result of frequent research and popularity of cryptocurrencies like Bitcoin and Ethereum. As a result, a blockchain is being considered a viable option in multiple application domains including government, finance, telecommunications and others. One of the major obstacles in the acceptability of blockchain is how it might perform in different application domains. Also, how different blockchain platforms would fair against each other in a number of criteria. This is particularly true for private blockchain platforms which require enterprise level performances. In order to address this issue, in this article, we have presented BlockMeter, an application agnostic performance measurement framework that enables to compare the performance of different private blockchain platforms, against some matrices, for any given application. We believe that this research will be a foundation for many performance evaluation tools. Such an evaluation tool could be an important component for many enterprises who are willing to integrate their applications with a blockchain platform. 

To show the applicability of our framework, we have evaluated the performance of two popular private blockchain platforms, Hyperledger Fabric and Hyperledger Sawtooth under different applications with different configuration parameters. From our experimental findings we have found Hyperledger Fabric and Hyperledger Sawtooth to be capable of handling complex computation and large data sets under comparatively lower traffic. However, under a very high request rate, Hyperledger Fabric has managed to deliver to some extent while Hyperledger Sawtooth has its performance degraded. This implicates that both of these platforms might require significant improvements to improve their performance. It is to be noted that all the experiments have been carried out under some default parameters (block-size: 10, batch timeout: 2s). These parameters could be tweaked to achieve better optimisation and improved performance \cite{thakkar2018performance}.

As of now, we have integrated support for two blockchain platforms from the Hyperledger projects. We intend to incorporate other private blockchain platforms from the Hyperledger project and beyond and experiment with them to understand how they would perform against each other. This will enable us or anyone interested to compare a wide range of private blockchain platforms for any given application. During the period of the current project and experiments, Hyperledger projects have also undergone several updates and it will be interesting to see how these updates have impacted their performance, if any.

